\documentclass[12pt,preprint]{aastex}

\shorttitle{Orbital Decay in PSR~B2127+11C}
\shortauthors{Jacoby et al.}

\begin{document}

\title{Measurement of Orbital Decay in the Double Neutron Star Binary PSR~B2127+11C}

\author{B.~A.~Jacoby\altaffilmark{1,2}, P.~B.~Cameron\altaffilmark{1}, F.~A.~Jenet\altaffilmark{3}, S.~B.~Anderson\altaffilmark{1}, R.~N.~Murty\altaffilmark{4}, and S.~R.~Kulkarni\altaffilmark{1}}

\altaffiltext{1}{Department of Astronomy, California Institute of
Technology, MS 105-24, Pasadena, CA 91125;
pbc@astro.caltech.edu, sba@astro.caltech.edu, srk@astro.caltech.edu.}

\altaffiltext{2}{present address: Naval Research Laboratory, Code 7213,
4555 Overlook Avenue, SW, Washington, DC 20375; bryan.jacoby@nrl.navy.mil.}

\altaffiltext{3}{Center for Gravitational Wave Astronomy, University of Texas at Brownsville, 80 Fort Brown, Brownsville, TX 78520; merlyn@alum.mit.edu}

\altaffiltext{4}{Cornell University, Ithaca, NY 14853; rnm5@cornell.edu}

\begin{abstract}
We report the direct measurement of orbital period decay in the double
neutron star pulsar system PSR~B2127+11C in the globular cluster M15
at the rate of $(-3.95\pm0.13)\times10^{-12}$, consistent with the
prediction of general relativity at the $\sim\!3 \%$ level.  We find
the pulsar mass to be $m_p = (1.358 \pm 0.010) M_\odot$ and the
companion mass $m_c = (1.354 \pm 0.010) M_\odot$.  We also report
long-term pulse timing results for the pulsars PSR~B2127+11A and PSR~B2127+11B, including confirmation of the cluster proper motion.
\end{abstract}


\section{INTRODUCTION}\label{sec:intro}

Pulsars in binary systems with neutron star companions provide the
best available laboratories for testing theories of gravity.  To date,
two such systems have been used for such tests: PSR~B1913+16
\citep{tw82, tw89}, and PSR~B1534+12 \citep{sac+98, sttw02}.  Both are
consistent with Einstein's general relativity (GR).

The globular cluster M15 (NGC~7078) contains 8 known radio pulsars,
the brightest of which are PSR~B2127+11A (hereafter M15A), a solitary
pulsar with a 110.6\,ms spin period; PSR~B2127+11B (M15B), a solitary
56.1\,ms pulsar; and PSR~B2127+11C (M15C), a 30.5\,ms pulsar in a
relativistic 8-hour orbit with another neutron star \citep{wkm+89,
and92}.  The Keplerian orbital parameters of M15C are nearly identical
to those of PSR~B1913+16, though the former did not follow the
standard high mass binary evolution \citep{agk+90}.  With our data set
spanning 12 years, M15C now provides a similar test of GR.

\section{OBSERVATIONS AND ANALYSIS}\label{sec:obs}

We observed M15 with the 305\,m Arecibo radio telescope from April
1989 to February 2001, with a gap in observations between February
1994 and December 1998 roughly corresponding to a major upgrade of the
telescope.  All observations used the 430\,MHz line feed, with 10\,MHz
of bandwidth centered on 430\,MHz.

Observations up to January 1994 were made with the Arecibo 3-level
autocorrelation spectrometer (XCOR), which provided 128 lags in each
of two circular polarizations and 506.625\,$\mu$s time resolution.
The autocorrelation functions were transformed to provide 128
frequency channels across the band, and these data were dedispersed
at the appropriate dispersion measure ($DM$) \citep{and92} and folded
synchronously with the pulse period for each pulsar.  Observations
were broken into sub-integrations of 10 minutes for M15A and M15C, and
20 minutes for the fainter M15B.

Beginning in January 1999, we used the Caltech Baseband Recorder (CBR)
for data acquisition.  This backend sampled the telescope signal in
quadrature with 2-bit resolution and wrote the raw voltage data to
tape for off-line analysis with the Hewlett-Packard Exemplar machine
at the Caltech Center for Advanced Computing Research (CACR).  After
unpacking the data and correcting for quantization effects
\citep{ja98}, we formed a virtual
32-channel filterbank in the topocentric reference frame with each
channel coherently dedispersed at the dispersion measure ($DM$) of
M15C \citep{hr75, jcpu97}.  The coherent filterbank data were then
dedispersed and folded for each pulsar as for the XCOR data.

\begin{figure}
\plotone{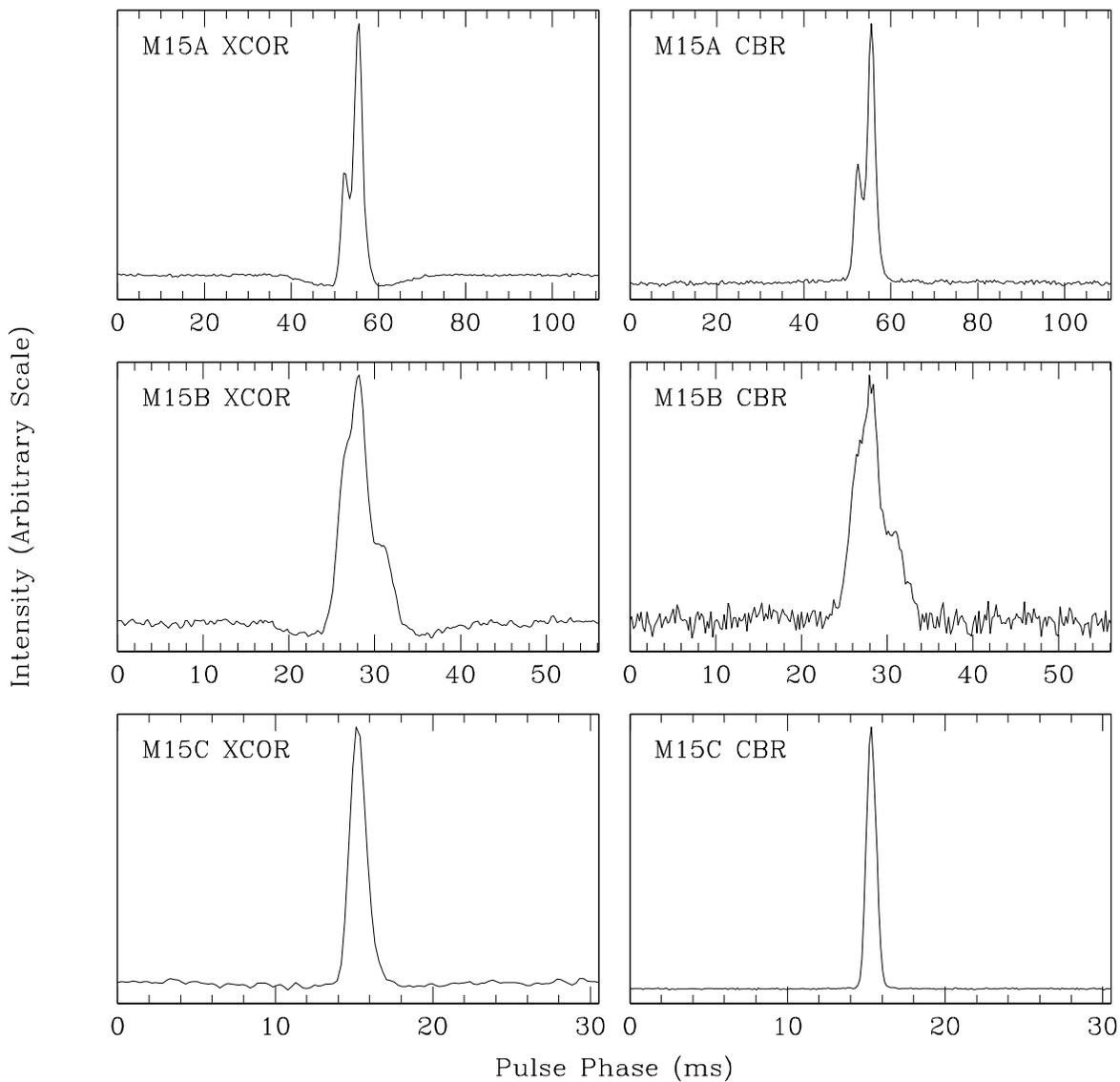}
\caption[Average pulse profiles for three pulsars in M15]{Average pulse profiles for three pulsars in M15.  Average pulse profiles of M15 A (top row), M15B (middle row), and M15C (bottom row) as observed with XCOR (left column) and CBR (right column).  The negative dips preceding and following the main pulse in the XCOR pulse profiles are an artifact of the correlator's coarse 3-level quantization; this distortion has been corrected in the CBR data \citep{ja98}.}
\label{fig:m15_profiles}
\end{figure}

The folded pulse profiles were cross-correlated against a high
signal-to-noise standard profile appropriate to the pulsar and backend
(Fig. \ref{fig:m15_profiles}) to obtain an average pulse time of
arrival (TOA) for each sub-integration, corrected to UTC(NIST).  The
standard pulsar timing package {\sc
tempo}\footnote{http://pulsar.princeton.edu/tempo}, along with the Jet
Propulsion Laboratory's DE405 ephemeris, was used for all timing
analysis.  TOA uncertainties estimated from the cross-correlation
process were multiplied by a constant determined for each
pulsar-instrument pair in order to obtain reduced $\chi^2 \simeq 1$.
An arbitrary offset was allowed between the XCOR and CBR data sets to
account for differences in instrumental delays and standard profiles.
The timing models resulting from our analysis are presented in Table
\ref{tab:par}, and post-fit TOA residuals relative to these models are
shown in Figure \ref{fig:m15_residuals}.  All stated and depicted
uncertainties correspond to 68\% confidence.

\section{DISCUSSION}\label{sec:discussion}

\subsection{Post-Keplerian Observables for M15C}\label{sec:pk}

In addition to the five usual Keplerian orbital parameters, in the
case of M15C we have measured three post-Keplerian (PK) parameters:
advance of periastron ($\dot\omega$), time dilation and gravitational
redshift ($\gamma$), and orbital period derivative ($\dot P_b$).  The
dependence of these PK parameters on the Keplerian parameters and
component masses depend on the theory of gravity; in GR, these
relations are (see Taylor \& Weisberg, 1982; Damour \& Deruelle, 1986; and Damour \& Taylor, 1992):\nocite{tw82,dd86,dt92}
\begin{eqnarray}
\dot\omega &=& 3G^{2/3}c^{-2}\left(\frac{P_{b}}{2 \pi}\right)^{-5/3} (1-e^2)^{-1} M^{2/3}\,, \label{eq:omdot}  \\
\gamma &=& G^{2/3}c^{-2}e\left(\frac{P_{b}}{2 \pi}\right)^{1/3} m_c(m_p + 2m_c) M^{-4/3}\,, \label{eq:gamma} \\
\dot P_{b} &=& -\,\frac{192\pi}{5}G^{5/3}c^{-5}\left(\frac{P_b}{2\pi}\right)^{-5/3}(1-e^2)^{-7/2} \left(1 + \frac{73}{24}e^2 + \frac{37}{96}e^4\right) m_p m_c M^{-1/3} \,,  \label{eq:pbdot} 
\end{eqnarray}
where $G$ is the gravitational constant and $c$ is the speed of light.
The measurement of any two PK observables determines the component
masses under the assumption that GR is the correct description of
gravity; measuring the third parameter overdetermines the system and
allows a consistency test of GR.

\subsection{Kinematic Effects on Pulse Timing}\label{sec:kin}

The rate of change of orbital period that we observe in the M15C system,
$(\dot{P}_b)^{\rm obs}$, is corrupted by kinematic effects that must
be removed to determine the intrinsic rate, $(\dot{P}_b)^{\rm int}$.
Following the discussion of \citet{phi92b, phi93} regarding the
parallel case of kinematic contributions to $\dot{P}$, we have

\begin{equation}\label{eq:kin}
\left(\frac{\dot{P_b}}{P_b}\right)^{\rm kin} = -\,\frac{v_0^2}{c R_0} \left(\cos b~\cos l + \frac{\delta - \cos b~\cos l}{1 + \delta^2 - 2\delta \cos b~\cos l}\right) + \frac{\mu^2 d}{c} - \frac{a_l}{c},
\end{equation}
where $v_0 = 220 \pm 20$\,km\,s$^{-1}$ is the Sun's galactic rotation
velocity, $R_0 = 7.7 \pm 0.7$\,kpc is the Sun's galactocentric
distance, $\delta \equiv d/R_0$, $\mu$ is the proper motion, $d =
9.98 \pm 0.47$\,kpc is the distance to the pulsar \citep{mhb04},
and $a_l$ is the pulsar's line-of-sight acceleration within the
cluster.  The first term in equation (\ref{eq:kin}) is due to the
pulsar's galactic orbital motion, the second to the secular
acceleration resulting from the pulsar's transverse velocity
\citep{shk70}, and the third to the cluster's gravitational field.

Acceleration within the cluster may well dominate the kinematic
contribution to $\dot{P}_b$, but $a_l$ is an odd function of the
distance from the plane of the sky containing the cluster center to
the pulsar, and since we do not know if M15C is in the nearer or
further half of the cluster, we must use its expectation value,
$\bar{a_l} = 0$.  \citet{phi93} calculates a maximum value of $\left|
a_l \right|_{\rm max} / c = 6 \times 10^{-18}$\,s$^{-1}$, too small
for the observed $\dot{P}$ to provide a useful constraint.  However,
the unknown $a_l$ still dominates the uncertainty of $(\dot{P}_b)^{\rm
kin}$; we take the median value of $0.71 \left| a_l \right| _{\rm
max}$ as the uncertainty in $a_l$
\citep{phi92b}.  Evaluating equation (\ref{eq:kin}), the total
kinematic contribution is

\begin{equation}
\left(\dot P_b\right)^{\rm kin} = (-0.0095\pm0.12)\times10^{-12}\,,
\end{equation}
and subtracting this contamination from $(\dot P_b)^{\rm obs}$ yields
the intrinsic value

\begin{equation}
\left(\dot P_b\right)^{\rm int} = (-3.95\pm0.13)\times10^{-12}\,.
\end{equation}

\subsection{Component Masses of M15C and a Test of General Relativity}\label{sec:mass}

\begin{figure}
\plotone{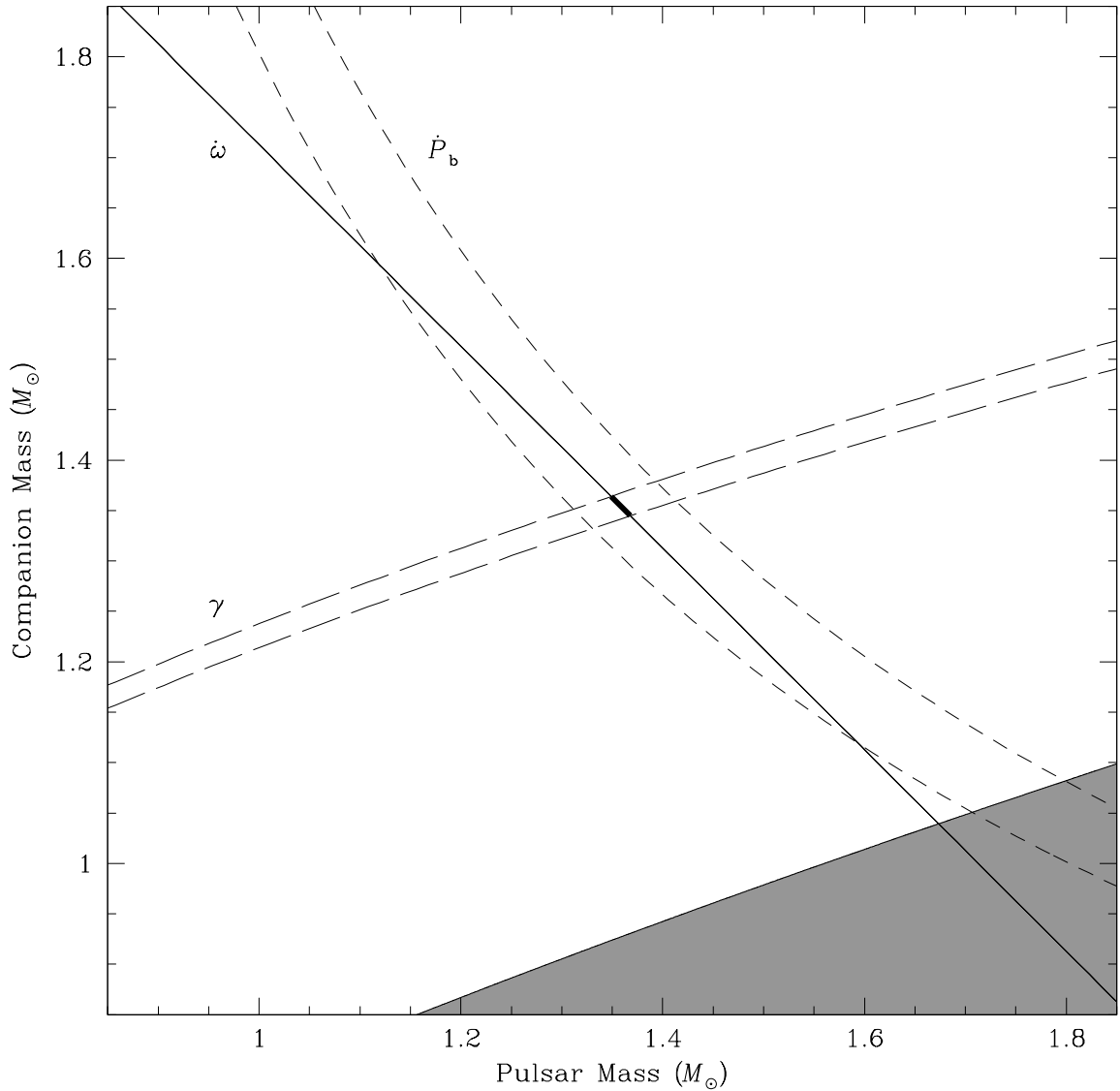}
\caption[M15C mass-mass diagram]{M15C mass-mass diagram.  The
constraints on the pulsar and companion masses in GR from the measured
values of $\gamma$ (long dashed lines), $\dot{\omega}$ (solid lines),
and intrinsic $\dot{P}_b$ (short dashed lines) are shown.  The allowed
region in mass-mass space at the intersection of these constraints is
denoted by a heavy line segment near the center of the plot.  The
shaded region is excluded by Kepler's laws.  The intersection of the
constraints from the three post-Keplerian observables indicates that
the behavior of this system is consistent with GR.}
\label{fig:m1m2GR}
\end{figure}

Solving equations (\ref{eq:omdot}) and (\ref{eq:gamma}) given the
measured values of $\dot{\omega}$, $\gamma$, $P_b$, and $e$ gives $m_p
= (1.358 \pm 0.010) M_\odot$, $m_c = (1.354 \pm 0.010) M_\odot$,
and $M
\equiv m_p + m_c = (2.71279 \pm 0.00013) M_\odot$ in the framework of
GR (Fig. \ref{fig:m1m2GR}).  This result is consistent with, and more
precise than, previous mass measurements for the neutron stars in the
M15C system \citep{pakw91, and92, dk96}.  We note that these masses
are consistent with the masses of double neutron star binaries observed
in the field \citep{tc99, sta04}.  M15 is a metal-poor cluster with a
mean metallicity [Fe/H] = -2.3 \citep{skp+91}, suggesting that the
mass of neutron stars is not a strong function of the metallicity of their
progenitors.

Our determination
of a third PK parameter gives a test of GR; $(\dot{P_b})^{\rm int}$ is
$1.003 \pm 0.033$ times the predicted value.  While M15C provides an
impressive test of GR, it is less stringent than the $1\%$
$\dot{\omega}$-$\gamma$-$\dot{P_b}$ test provided by PSR~B1913+16
\citep{tw89} and the $0.5\%$ $\dot{\omega}$-$\gamma$-$s$ test provided
by PSR~B1534+12
\citep{sttw02}, where $s \equiv \sin i$ is the shape parameter determined through measurement of Shapiro delay.  We note that the uncertainty in the intrinsic orbital
period decay is due almost entirely to the kinematic contribution, so
further observations will not significantly improve our determination
of $(\dot{P}_b)^{\rm int}$ or the quality of the test of GR it allows.

\subsection{Proper Motion of M15}\label{sec:pm}

\begin{figure}
\plotone{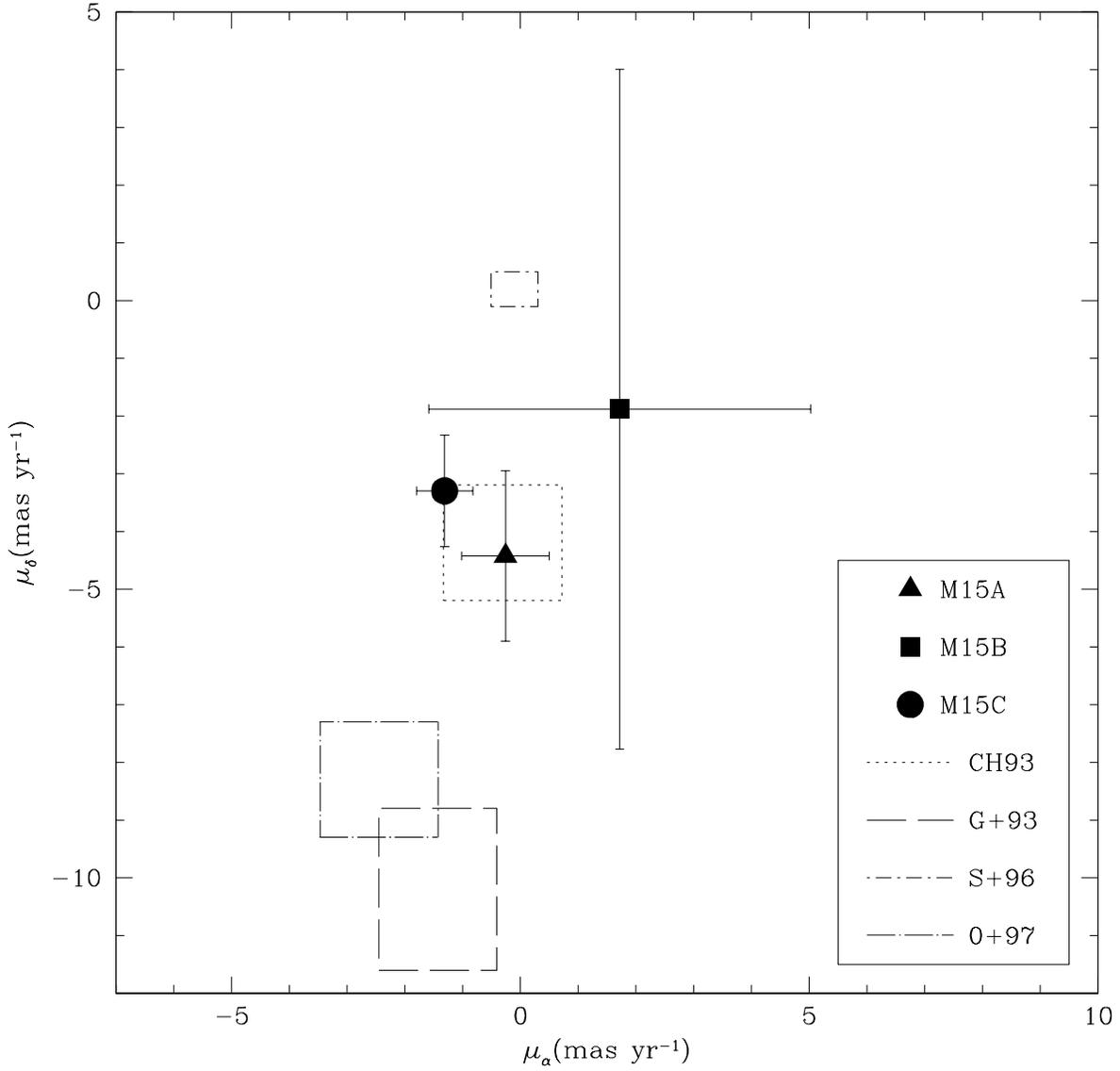}
\caption[Proper motion of M15]{Proper motion of M15.  Measured proper
motions of M15A, M15B, and M15C in right ascension and declination,
with rectangular regions indicating the published cluster proper
motion measurements of \citet{ch93} (CH93), \citet{gcl+93} (G+93),
\citet{soh+96} (S+96), and \citet{obg+97} (O+97).}
\label{fig:m15_pm}
\end{figure}

The proper motions resulting from our timing analysis give absolute
transverse velocities for M15A and M15C several times greater than the
cluster escape velocity.  The measured proper motions for these two
pulsars and M15B are shown in Figure \ref{fig:m15_pm}, along with four
published proper motion measurements for M15 based on optical
astrometry.  The pulsar proper motions are all consistent with each
other; their average is $\mu_{\alpha} = (-1.0 \pm 0.4)\,{\rm
mas\,yr}^{-1}$, $\mu_{\delta} = (-3.6 \pm 0.8)\,{\rm mas\,yr}^{-1}$.
This result is in good agreement with the cluster measurement of \citet{ch93}.

\subsection{Intrinsic Spin Period Derivatives}

If we assume that GR provides the correct description of gravity, we
can use $(\dot{P_b})^{\rm obs}$ to determine the total kinematic
correction to $\dot{P_b}$ and hence, to $\dot{P}$ for M15C.  We find 

\begin{equation}\label{eq:pbdotkin}
\left(\frac{\dot{P_b}}{P_b}\right)^{\rm kin}_{\rm GR} = \left(\frac{\dot{P}}{P}\right)^{\rm kin}_{\rm GR} = \left( -8 \pm 17 \right) \times 10 ^{-19} \,{\rm s}^{-1}\,, 
\end{equation}
which corresponds to $a_l / c = \left( 4 \pm 17 \right) \times 10
^{-19} \, {\rm s}^{-1}$.  We now apply this correction to the observed
value of $\dot{P}$ and find the intrinsic value assuming GR,
$(\dot{P})^{\rm int}_{\rm GR} = (0.00501 \pm 0.00005) \times
10^{-15}$.  This intrinsic spindown rate allows us to improve upon the
previous estimate of the pulsar's characteristic age and magnetic
field strength \citep{and92}; we find $\tau_c = (0.097 \pm
0.001)$\,Gyr and $B_{\rm surf} = (1.237 \pm 0.006) \times 10^{10}$\,G.

\begin{figure}
\plotone{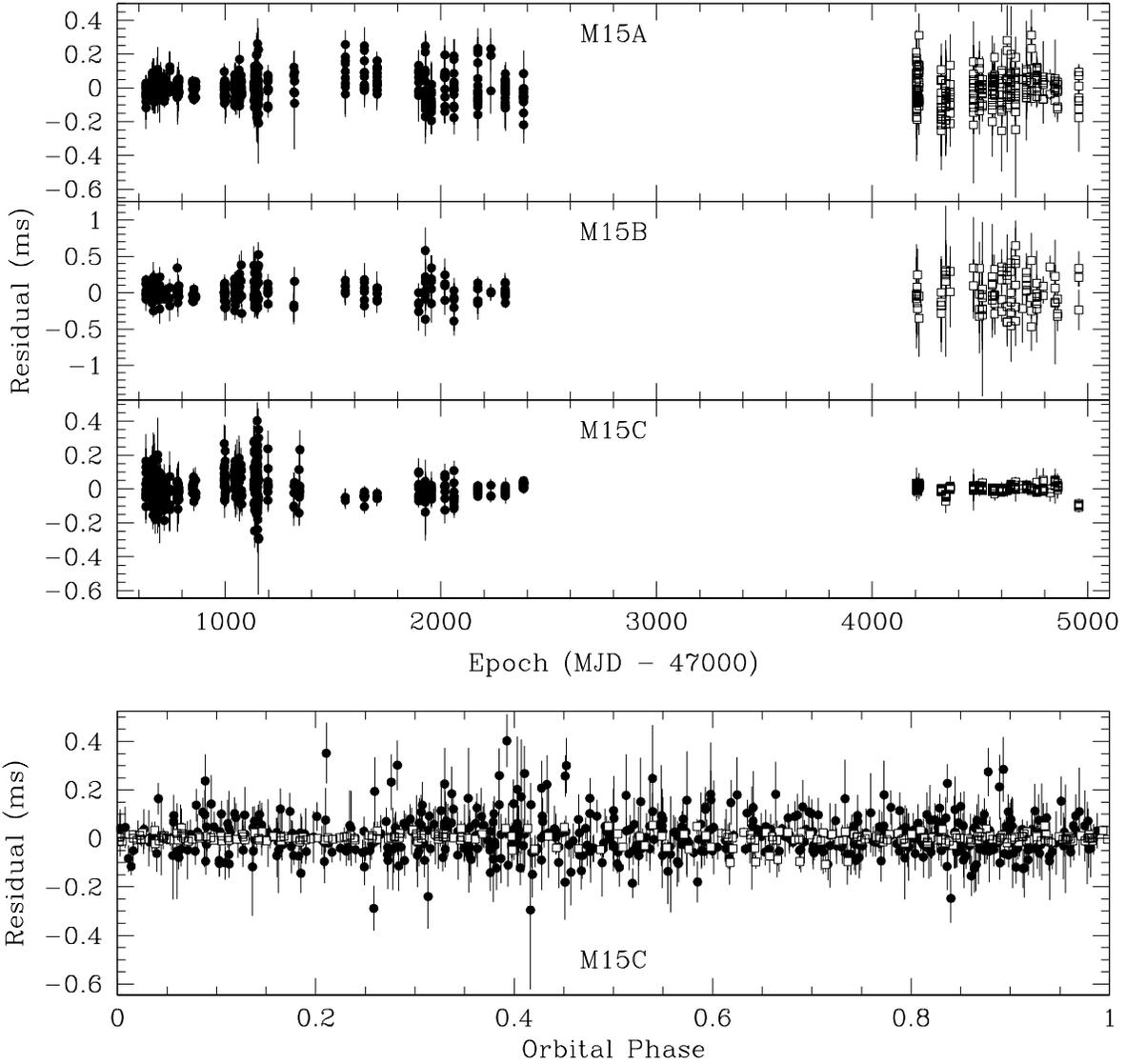}
\caption[Timing residuals for three pulsars in M15]{Timing residuals
of M15A, M15B, and M15C.  Residuals for each pulsar versus observation
epoch are shown in the top three panels, with M15C residuals versus
orbital phase in the bottom panel.  Filled circles represent XCOR
observations, with open squares indicating CBR data.}
\label{fig:m15_residuals}
\end{figure}

Our timing models for M15A and M15C include $\ddot{P}$
(Tab. \ref{tab:par}) which is unlikely to be intrinsic to the
pulsars.  For M15A in the cluster core, \citet{phi93} estimates the
kinematic contribution to be $\left| \dot{a_l} / c \right| \equiv \left|
\ddot{P} / P \right| \le 10^{-26}$\,s$^{-1}$ (80\% confidence).  This is 
significantly larger than the observed $\left| \ddot{P} / P \right| \sim 3 
\times 10^{-28}$\,s$^{-1}$, so the observed $\ddot{P}$ is consistent with the
expected jerk resulting from the cluster's mean field and nearby
stars.  For M15C, far from the cluster core, we measure $\left|
\ddot{P} / P \right| \sim 10^{-28}$\,s$^{-1}$.  We note that our measurement 
of $\ddot{P}$ in M15C is not of high significance ($\sim\!2 \sigma$), and may 
be an artifact of the systematic trends apparent in our timing data (Fig. \ref{fig:m15_residuals}).

\acknowledgments

The Arecibo Observatory, a facility of the National Astronomy and
Ionosphere Center, is operated by Cornell University under cooperative
agreement with the National Science Foundation.  We thank W. Deich for
providing the pulsar data analysis package, {\sc psrpack}.  BAJ and
SRK thank NSF and NASA for supporting this research.  Part of this
research was carried out at the Jet Propulsion Laboratory, California
Institute of Technology, under a contract with the National
Aeronautics and Space Administration and funded through the internal
Research and Technology Development program.  BAJ holds a National
Research Council Research Associateship Award at the Naval Research
Laboratory (NRL).  Basic research in radio astronomy at NRL is
supported by the Office of Naval Research.

\begin{deluxetable}{lrrr}
\tablecaption{Pulsar Parameters for B2127+11A--C\label{tab:par}}
\tablecolumns{4}
\tabletypesize{\scriptsize}
\tablewidth{0pt}
\tablehead{
    \colhead{Parameter\tablenotemark{a}} 
      & \multicolumn{3}{c}{Pulsar}
\\
\cline{2-4} \\
      & \colhead{B2127+11A}
      & \colhead{B2127+11B}
      & \colhead{B2127+11C}
}
\startdata
Right ascension, $\alpha_{\rm J2000}$\dotfill
   & $21^{\rm h}29^{\rm m}58\fs2472(3)$
   & $21^{\rm h}29^{\rm m}58\fs632(1)$
   & $21^{\rm h}30^{\rm m}01\fs2042(1)$
\\
Declination, $\delta_{\rm J2000}$\dotfill
   & $+12\arcdeg10\arcmin01\farcs264(8)$
   & $+12\arcdeg10\arcmin00\farcs31(3)$
   & $+12\arcdeg10\arcmin38\farcs209(4)$
\\
Proper motion in $\alpha$, $\mu_{\alpha}$ (mas yr$^{-1}$)\dotfill
   & -0.26(76)
   & 1.7(33)
   & -1.3(5)
\\
Proper motion in $\delta$, $\mu_{\delta}$ (mas yr$^{-1}$)\dotfill
   & -4.4(15)
   & -1.9(59)
   & -3.3(10)
\\Pulse period, $P$ (ms)\dotfill
   & 110.66470446904(5)
   & 0.05613303552473(9)
   & 30.52929614864(1)
\\
Reference epoch (MJD)\dotfill
    & 50000.0
    & 50000.0
    & 50000.0
\\
Period derivative, $\dot{P}$ (10$^{-15}$)\dotfill
   & -0.0210281(2)
   & 0.0095406(6)
   & 0.00498789(2)
\\
Period second derivative, $\ddot{P}$ (10$^{-30}$\,s$^{-1}$)\dotfill
   & 32(5)
   & \nodata
   & -2.7(13)
\\
Dispersion measure, $DM$ (pc cm$^{-3}$)\dotfill
   & 67.31
   & 67.69
   & 67.12
\\
Binary model\dotfill
   & \nodata
   & \nodata
   & DD
\\
Orbital period, $P_b$ (d)\dotfill
   & \nodata
   & \nodata
   & 0.33528204828(5)
\\
Projected semimajor axis, $a \sin i$ (lt-s)\dotfill
   & \nodata
   & \nodata
   & 2.51845(6)
\\
Orbital eccentricity, $e$\dotfill
   & \nodata
   & \nodata
   & 0.681395(2)
\\
Longitude of periastron, $\omega$ (deg)\dotfill
   & \nodata
   & \nodata
   & 345.3069(5)
\\
Time of periastron, $T_0$\dotfill
   & \nodata
   & \nodata
   & 50000.0643452(3)
\\
Advance of periastron, $\dot{\omega}$ (deg\,yr$^{-1}$)\dotfill
   & \nodata
   & \nodata
   & 4.4644(1)
\\
Time dilation \& gravitational redshift, $\gamma$ (s)\dotfill
   & \nodata
   & \nodata
   & 0.00478(4)
\\
Orbital period derivative, $(\dot{P_b})^{\rm obs}$ ($10^{-12}$)\dotfill
   & \nodata
   & \nodata
   & -3.96(5)
\\
Weighted RMS timing residual ($\mu$s)\dotfill &
58.9 & 103.5 & 26.0
\\
\cutinhead{Derived Parameters}
Orbital period derivative, $(\dot{P_b})^{\rm int}$ ($10^{-12}$)\dotfill
   & \nodata
   & \nodata
   & -3.95(13)\tablenotemark{b}
\\
Pulsar mass, $m_p$ ($M_\odot$)\dotfill
   & \nodata
   & \nodata
   & 1.358(10)
\\
Companion mass, $m_c$ ($M_\odot$)\dotfill
   & \nodata
   & \nodata
   & 1.354(10)
\\
Total mass, $M = m_p + m_c$ ($M_\odot$)\dotfill
   & \nodata
   & \nodata
   & 2.71279(13)
\\
Galactic longitude, $l$ (deg)\dotfill
    & 65.01
    & 65.01
    & 65.03
\\
Galactic latitude, $b$ (deg)\dotfill
    & -27.31
    & -27.31
    & -27.32
\\
Transverse velocity, $v_{\perp}$ (km s$^{-1}$)\tablenotemark{c}\dotfill
    & 210(70)
    & 120(230)
    & 170(40)
\\
Intrinsic period derivative, $\dot{P}_{\rm int}$ (10$^{-15}$)\tablenotemark{b}\dotfill 
   & \nodata
   & \nodata
   & 0.00501(5)
\\
Surface magnetic field, $B_{\rm surf}$ $(\times 10^10 \rm{G})$\tablenotemark{b}\dotfill 
   & \nodata
   & \nodata
   & 1.237(6)
\\
Characteristic age, $\tau_c$ (Gyr)\tablenotemark{b}\dotfill 
   & \nodata
   & \nodata
   & 0.097(1)
\\

\enddata
\tablenotetext{a}{Figures in parenthesis are $1\,\sigma$ (68\%) uncertainties in the last
digit quoted.  Uncertainties are calculated from twice the formal
error produced by {\sc tempo}.}
\tablenotetext{b}{Corrected for kinematic effects as described in \S \ref{sec:kin}}
\tablenotetext{c}{Based on the distance measurement of \citet{mhb04}}
\end{deluxetable}

\end{document}